\documentstyle[12pt]{article}
\begin{document}
\font\twelve=cmbx10 at 15pt
\font\ten=cmbx10 at 12pt
\font\eight=cmr8

\begin{titlepage}

\begin{center}

{\ten Centre de Physique Th\'eorique\footnote{Unit\'e Propre de
Recherche 7061} - CNRS - Luminy, Case 907}

{\ten F-13288 Marseille Cedex 9 - France }

\vspace{1 cm}

{\twelve Critical Properties of Stochastic Cellular Automata}

\vspace{0.3 cm}

{\bf Stam NICOLIS\footnote{Centre de Physique Th\'eorique, CNRS,
Luminy, Case 907, F-13288 Marseille cedex 9, France {\em and}
PHYMAT, Universit\'e de Toulon et du Var, 83957 La Garde cedex, France}}

\vspace{1.5 cm}

{\bf Abstract}

\end{center}
We study the effect of mixing two rules on the dynamics of
one-dimensional cellular automata by large scale numerical
simulations. We calculate the decay of the  
magnetization
 for the Domany-Kinzel 
automaton (XOR/AND mixing) 
to its equilibrium value in the three phases. This requires system
sizes in excess of 1 million sites. We also find severe finite size
effects near the new critical points recently proposed on the basis of
transfer matrix arguments.

\vspace{2 cm}


\bigskip

\noindent Number of figures : 4

\bigskip

\noindent May 1994

\noindent CPT-94/P.3037

\noindent cond-mat/9406005

\bigskip

\noindent anonymous ftp or gopher: cpt.univ-mrs.fr

\begin{tabular}{lc}
World Wide Web: & http://cpt.univ-mrs.fr {\em or} \\
                & gopher://cpt.univ-mrs.fr/11/preprints\\
\end{tabular}

\end{titlepage}

\begin{titlepage}

\begin{center}
{\Large\bf Critical Properties of Stochastic Cellular Automata}

\vskip 1truecm
Stam Nicolis

\vskip 1truecm
{\sl Centre de Physique Th\'eorique, CNRS-Luminy, Case 907\\
F-13288 Marseille cedex 9, France}

\vskip 1truecm
ABSTRACT 

\end{center}
We study the effect of mixing two rules on the dynamics of
one-dimensional cellular automata by large scale numerical
simulations. We calculate the decay of the  
magnetization
 for the Domany-Kinzel 
automaton (XOR/AND mixing) 
to its equilibrium value in the three phases. This requires system
sizes in excess of 1 million sites. We also find severe finite size
effects near the new critical points recently proposed on the basis of
transfer matrix arguments.

\end{titlepage}
Cellular automata have been studied in a systematic way for more than
ten years~\cite{Wolfram} and constitute a deceptively simple family of
dynamical systems-where space, time and the range of values of the
dynamical variable have been discretized; yet still there are many
open problems. Even better, one finds new features in well studied
systems. 

As a concrete illustration we shall study the Domany-Kinzel
automaton~\cite{DoKi}, defined as follows:
\begin{equation}
\label{dk}
\begin{array}{c}
P(\sigma_{i}^{t+1}=1|\sigma_{i-1}^{t}=1,\sigma_{i+1}^{t}=1)=p_2\\
P(\sigma_{i}^{t+1}=1|\sigma_{i-1}^{t}=1,\sigma_{i+1}^{t}=0)=
P(\sigma_{i}^{t+1}=1|\sigma_{i-1}^{t}=0,\sigma_{i+1}^{t}=1)=p_1\\
P(0|1,0)=1-p_1;\,\,P(0|1,1)=1-p_2;\,\,P(1|0,0)=0\\
\end{array}
\end{equation}
starting from some initial configuration
$\{\sigma_{i}^{t=0}\}_{i=1}^{N}$.  If $p_1=1$ and $p_2=0$ one readily
identifies the XOR rule, $P(1|1,0)=P(1|0,1)=1,\,\,P(1|1,1)=0$, while,
if $p_1=0$ and $p_2=1$ one has the AND rule, $P(1|1,1)=1$ all others
giving zero. 

This system has been the object of many investigations since its
introduction in ref~\cite{DoKi}-\cite{Ki,Tsallis,KohSchr,BaLiRu}, since many
models of interest may be mapped onto it, in some region of its
parameter space. Nevertheless many open questions remain, first among
them being the accurate determination of the transition points and a
study of its critical properties~\cite{ZePe}.
Whereas the original papers~\cite{DoKi,Ki} identified one critical point (from
a mapping to an exactly solvable model), namely at $p_1=1,p_2=0.5$ and
conjectured the existence of another one (but could not resolve the
issue due to the small system size), recent work has not only--essentially-- settled
this issue (in the positive sense)~\cite{Tsallis,KohSchr,ZePe}, but has
advanced the conjecture of the existence of novel critical points
within the active and the chaotic phases~\cite{BaLiRu}. 

In the present note we present results from large scale simulations on
the decay of the magnetization in the various phases of the model. 
We also investigate what happens near the (critical) points
$(p_1=0.614,p_2=0.925)$,and $(p_1=0.648,p_2=0.861)$,
conjectured to exist on the basis of approximate renormalization group
calculations~\cite{BaLiRu}. 

\begin{figure}
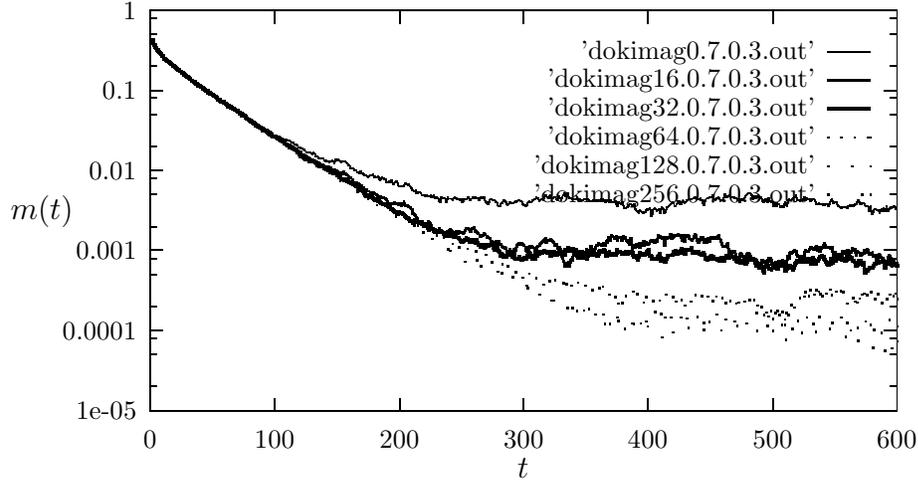

\input mag4k-1M.tex
\caption[]{Magnetization vs. time in the frozen
phase--$(p_1=0.3,p_2=0.7)$ for several
sizes--4096,65536,131072,262144,524288,1048576 from top to bottom.}
\label{size}
\end{figure}

First of all one finds that the system sizes studied to date (even
those of ref.~\cite{KohSchr}) are too small to resolve this sort
of question. As an example we display in fig.~\ref{size} the
magnetization, $m(t)$ against time for system sizes up to 1Msites.
Especially in the active and chaotic phases, equilibrium is reached so fast that
it is impossible to measure how the magnetization decays to its
equilibrium value. The breakpoint (in the frozen phase) appears to be 128ksites. Therefore
all our runs were henceforth done at 128ksites and larger. In fact,
even this size was too small to resolve many situations and 4 Msites
was barely satisfactory (see below). On the other hand,  
In the
frozen phase we find exponential decay with characteristic time
$\tau\approx 80$, fig.~\ref{size}. 

In the active and chaotic phases, we find exponential decay, with a
characteristic time essentially equal to that in the frozen phase for
the active phase; the chaotic phase, on the other hand, has a
significantly longer decay time, as might be expected.  

\begin{figure}
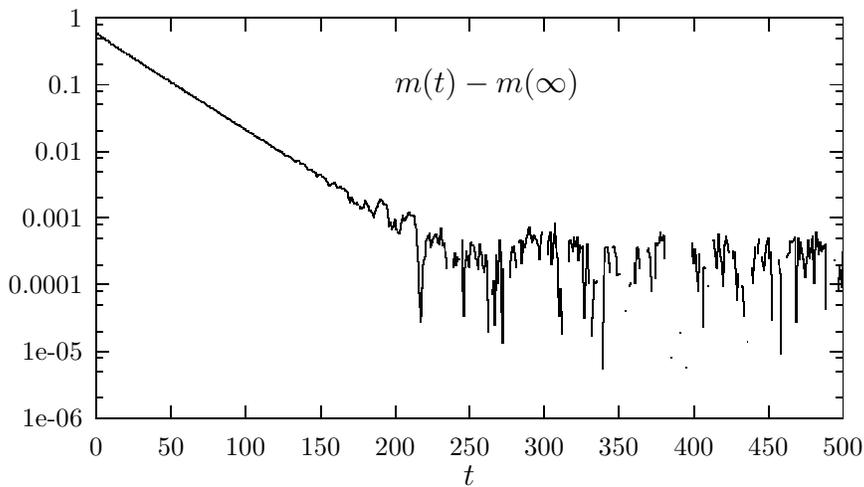

\input mag-4M.a.tex
\caption[]{Magnetization vs. time in the active
phase--$(p_1=0.75,p_2=0.87)$ for 4194304 sites.}
\end{figure}
\vfill\eject
\begin{figure}
\input mag-4M.c.tex
\caption[]{Magnetization vs. time just inside the chaotic
phase--$(p_1=0.76,p_2=0.2)$ for 4194304 sites and two different initial
conditions.}
\end{figure}
\begin{figure}[tbp]
\vskip -3truecm
\special{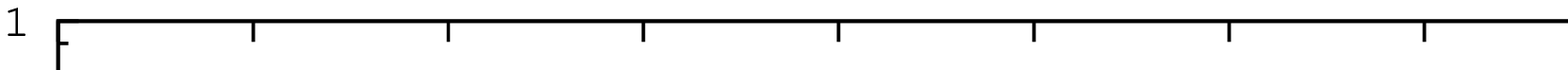 hscale=50
vscale=50 }
\vskip 10truecm
\label{RH1}
\caption[]{Normalized magnetization,$M(t)=m(t)-m(\infty)$ vs. time for
the point $(p_1=0.641,p_2=0.861)$
(point $R$) and the point $(p_1=0.614,p_2=0.925)$ (point $H_1$);
$L=131072$. }
\end{figure}
Regarding what happens at the points $R$ and $H_1$, we find 
that the values given for $p_1,p_2$ are almost surely misleading--the
finite time effects from the transfer matrix method must be much too
strong. 
For instance, for  $L=131072$ we find an
exponential decay with definitely different decay times--cf.
Fig.~\ref{RH1}. 

To summarize, we have studied the time dependence of the magnetization
in the three phases of the Domany-Kinzel cellular automaton. 
Our results show
that 4 Msites barely allow one to reach the asymptotic regime slowly
enough to extract a relaxation time. Such large systems and long times
are needed to
resolve the differences (if they do exist) between the {\em true}
points $H_1$ and $R$. On the other hand it should be noted that severe
critical slowing down near these critical points is a problem at these
sizes, as preliminary simulations have indicated.      
Even larger sizes are
necessary to study the  properties of higher correlation
functions~\cite{Stam}-however
these systems are definitely within reach of contemporary
supercomputers and even fast workstations (memorywise there is no
problem in running even 16 Msite systems on a Sparc 10 (or even an
ELC) workstation and
on a Cray YMP memory and time requirements fall within the ``tiny''
class). These large sizes and long times are necessary; since already,
for much smaller sizes, we find, for
instance, non-zero damage at $p_1=1,p_2=0.38$ (for $L=65536$ and $1\%$
initial damage)  and for 
$p_1=0.74,p_2=0.0$,(for $L=45056$ and $1\%$ initial damage)--cf.
ref.~\cite{ZePe}.  

{\bf Acknowledgements}: the code was developed and the first results
were obtained on the Cray Y-MP of
the J\"ulich Research Center; later results were obtained on the Cray
Y-MP of the Mediterranean
Institute of Technology. Computer time on these machines is gratefully
acknowledged. I would also like to thank G. A. Kohring, R. Livi, 
H. J. Herrmann and D. Stauffer for stimulating discussions.

\end{document}